\documentclass[fleqn,usenatbib]{mnras}

\usepackage[T1]{fontenc}
\usepackage[table]{xcolor}  

\DeclareRobustCommand{\VAN}[3]{#2}
\let\VANthebibliography\thebibliography
\def\thebibliography{\DeclareRobustCommand{\VAN}[3]{##3}\VANthebibliography}

\usepackage{graphicx}	
\usepackage{amssymb}
\usepackage{amsmath}	
\usepackage{graphicx}
\usepackage{rotating}
\usepackage{capt-of}
\usepackage{adjustbox}
\usepackage{xspace}

\newcommand{\fluxcgs}{erg~s$^{-1}$~cm$^{-2}$\xspace}

\newcommand{\nicer}{\textit{NICER}\xspace}

\newcommand{\src}{4U~1608--522\xspace}

\title[The 2020 Superburst of \src]{The 2020 Superburst of \src~and its impact on the accretion disk}

\author[Boztepe, T. et al.]
{Tu\u{g}ba Boztepe$^{1},$ 
Tolga G\"uver$^{2,3,4}$,
Elif Ece Devecio\u{g}lu$^{1}$,
Julia Speicher$^{4,5}$,
Motoko Serino$^{6}$,
\newauthor
David R. Ballantyne$^{4,5}$,
Diego Altamirano$^{7}$,
Gaurava K. Jaisawal$^{8}$,
Mason Ng$^{9,10}$,
\newauthor 
Andrea Sanna$^{11}$
Can Güngör$^{2,3}$,
Wataru Iwakiri$^{12}$
\\
\\
$^{1}$Istanbul University, Graduate School of Sciences, Department of Astronomy and Space Sciences, Beyazıt, 34119, İstanbul, T\"{u}rkiye \\
$^{2}$Istanbul University, Science Faculty, Department of Astronomy and Space Sciences, Beyaz{\i}t, 34119, \.{I}stanbul, T\"{u}rkiye \\
$^{3}$Istanbul University Observatory Research and Application Center, Istanbul University 34119, \.{I}stanbul, T\"{u}rkiye \\
$^{4}$School of Physics, Georgia Institute of Technology, Atlanta, 30332, USA\\
$^{5}$Center for Relativistic Astrophysics, School of Physics, Georgia Institute of Technology, 837 State Street, Atlanta, GA 30332-0430, USA\\
$^{6}$Department of Physical Science, Aoyama Gakuin University, 5-10-1 Fuchinobe, Chuo-ku, Sagamihara, Kanagawa 252-5258, Japan \\
$^{7}$School of Physics and Astronomy, University of Southampton, Southampton SO17 1BJ, UK\\
$^{8}$DTU Space, Technical University of Denmark, Elektrovej 327-328, DK-2800 Lyngby, Denmark\\
$^{9}$Department of Physics, McGill University, 3600 rue University, Montr\'{e}al, QC H3A 2T8, Canada\\
$^{10}$Trottier Space Institute, McGill University, 3550 rue University, Montr\'{e}al, QC H3A 2A7, Canada\\
$^{11}$Dipartimento di Fisica, Universit\`a degli Studi di Cagliari, SP Monserrato-Sestu, KM 0.7, Monserrato, 09042 Italy \\
$^{12}$International Center for Hadron Astrophysics, Chiba University, 1-33 Yayoi, Inage, Chiba 263-8522, Japan \\
\date{Accepted XXX. Received YYY; in original form ZZZ}
}
\pubyear{\the\year{}}

\begin{document}
\label{firstpage}
\pagerange{\pageref{firstpage}--\pageref{lastpage}}
\maketitle

\begin{abstract}

Superbursts are rare events observed from bursting neutron star low mass X-ray binaries. They are thought to originate from unstable burning of the thick layer of Carbon on the surface of the neutron star, causing the observed X-ray flashes to last several hours. Given their fluence it has long been thought that superbursts may have significant effects on the accretion flow around the neutron star. In this paper, we first present evidence for a new superburst observed from \src by MAXI during the 2020 outburst, around 00:45 UTC on 16 July 2020. We compare some of the properties of this superburst and the underlying outburst with the events recorded on May 5 2005 by RXTE and most recently in 2025 by MAXI. We then present our spectral analysis of \nicer and Insight-HXMT data obtained before and after the 2020 superburst event. Our results indicate that the inner disk temperature and the radius show a systematic evolution in the following few days, which may be related to the superburst. We show that the timescale of the observed evolution can not be governed by viscous timescales unless the viscosity parameter is unrealistically low.

\end{abstract}

\section{Introduction}
Low-mass X-ray binaries (LMXBs) are systems consisting of a compact object -- either a neutron star~(NS) or a black hole~(BS) -- and a companion star of low mass, typically of late spectral type. In some neutron star LMXBs, sudden X-ray flashes, known as thermonuclear bursts, occur due to unstable nuclear burning of accreted material on the neutron star's surface. These bursts typically exhibit blackbody spectra, and their flux profiles are described by a power-law~(herafter POW) decay \citep[see e.g.,][]{Cumming_2004,Cumming_2006a,2006csxs.book..113S,2014A&A...562A..16I,2017A&A...606A.130I}.

The composition of the accreted material significantly influences burst profiles. Hydrogen-rich accretion typically leads to longer bursts due to the rp-process, while helium-rich accretion generally results in shorter, more intense bursts, although intermediate-duration helium bursts can last up to an hour, particularly on relatively cool neutron stars. Superbursts are a distinct class of X-ray bursts characterized by their exceptional duration, lasting several hours, and are believed to result from the unstable burning of carbon in deeper layers of the neutron star’s accreted envelope \citep{Cumming_2001,Strohmayer_2002,Keek_2011}. These events are much rarer and about $10^{3}$ times more energetic than typical Type I X-ray bursts, provide insights into the long-term nuclear processes occurring on neutron stars \citep{Keek_2008,2017symm.conf..121I}.

Superbursts have been observed in several sources, including~4U~1735--44~\citep{2000A&A...357L..21C}, KS~1731--260~\citep{Kuulkers_2002}, GX~3$+$1~\citep{Kuulkers_2002b}, 4U~1820--30~\citep{Strohmayer_2002,2025arXiv250407329J},  4U~1636--536 
\citep{2001ApJ...554L..59W} and Aql~X--1~\citep{2021ApJ...920...35L}. To date, 28 superbursts have been observed from 16 different sources \citep{10.1093/mnras/stad374}. Most recently, a new superburst was observed from \src on March 19, 2025, by MAXI, lasting at least 3 hours \citep{Serino_2025,2025ATel17118....1C}.

\src, is a transient NS LMXB first discovered in 1971 by two Vela-5 satellites \citep{Tananbaum_1976,Belian_1976}. The system comprises a neutron star and a hydrogen-rich companion star, as indicated by a 12.9~hour orbital period observed in the optical band \citep{Wachter_2002,Keek_2008}. The distance of \src is estimated with values ranging from a lower limit of 3.0~kpc to approximately 4.1~kpc \citep{Güver_2010,2016ApJ...820...28O}. 

Based on data from Rossi X-ray Timing Explorer (RXTE), BeppoSAX, and INTEGRAL archives, the Multi-INstrument Burst ARchive \citep[MINBAR, ][]{2020ApJS..249...32G}, catalog contains 147 bursts detected from \src. The first superburst was detected on May 5, 2005, with RXTE All Sky Monitor (ASM) and High Energy Transient Explorer (HETE) data \citep{Keek_2008}. This event was the first superburst observed from a “classical” transient source, unlike previously known superburst sources that showed persistent accretion~\citep{Keek_2008}. Despite this difference in accretion behavior, the properties of the \src superburst were broadly comparable to those of other superbursts \citep{Keek_2008}. 

In this study, we report the detection of a new superburst from \src during its 2020 outburst, and investigate its possible impact on the accretion disk. The paper is organized as follows. Section~\ref{sec:superburst} provides an overview of the 2020 outburst of and shows spectral evidence for a superburst from \src in 2020. Section~\ref{sec:analaysis} presents the pointed observations of \src obtained by \nicer and HXMT around the 2020 superburst, including details on data reduction, spectral modeling, and the evolution of spectral parameters. Section~\ref{sec:discussion} discusses the implications of our findings, with subsections focusing on the comparison with previous outbursts, superburst characteristics, and the evolution of disk accretion. Finally, Section~\ref{sec:conclusions} summarizes our results and conclusions.

\section{2020 Outburst of \src~and the superburst} 
\label{sec:superburst}

\begin{figure*}
\includegraphics[scale=0.6]{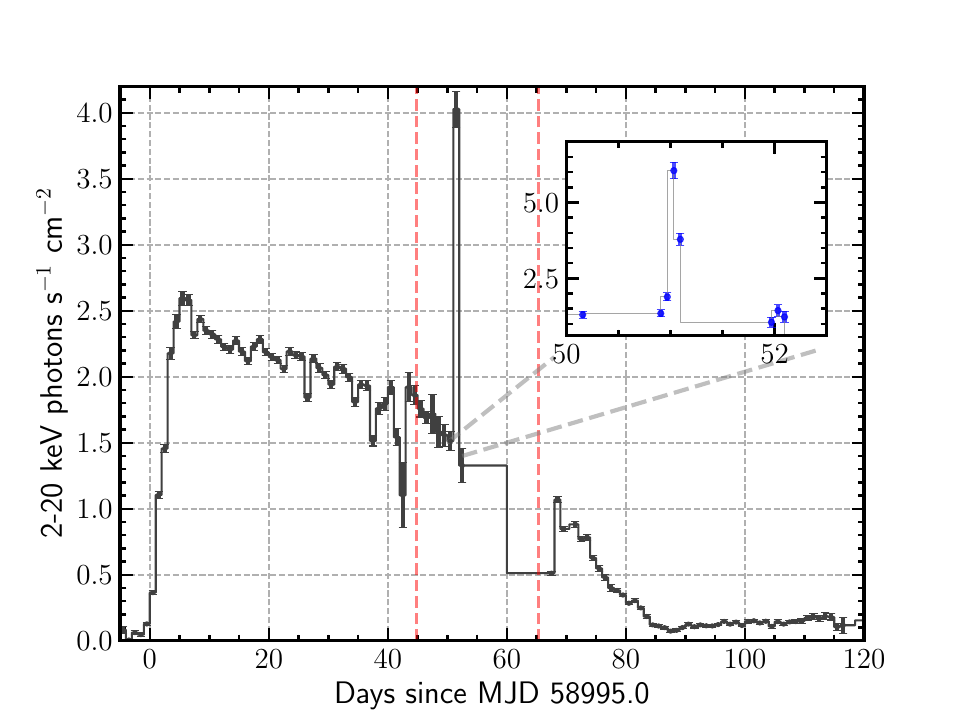}
\centering
\caption{2--20~keV light curve of the 2020 outburst of \src observed by MAXI. While the main plot shows daily average count rates, the inset shows the per orbital data obtained from MAXI around the superburst times. The time interval covered by the \nicer and HXMT observations are shown with red vertical dashed lines.} 
\label{maxi_lc}
\end{figure*}

The Monitor of All-sky X-ray Image (MAXI; \citealt{Matsuoka_2009}) is an all-sky monitor following the long-term intensity variations of X-ray sources, since August 2009. \src, exhibited an increase in X-ray flux in May 2020, marking the onset of a new outburst~(see \autoref{maxi_lc}). Within approximately 5 days, the source flux increased from $\sim$0.1 photons~s$^{-1}$cm$^{-2}$ to $\sim$2.5 photons~s$^{-1}$cm$^{-2}$ at the peak of the outburst in 2--20~keV range. The outburst then gradually decayed over a period of about 80 days, eventually reaching the low-level activity level. \autoref{maxi_lc} presents the 2--20~keV MAXI light curve of \src during the 2020 outburst.

On the 51st day of the outburst~(July 16, 2020, MJD 59046.03125), the MAXI flux of the source increased from about 1.35~photons~s$^{-1}$cm$^{-2}$ to 6~photons~s$^{-1}$cm$^{-2}$ in the 2--20~keV range, then the next orbit (90 minutes after) decayed to about 3.78~photons~s$^{-1}$cm$^{-2}$~(see \autoref{maxi_lc}).

\begin{figure*}
     \centering 
    \includegraphics[scale=0.54]{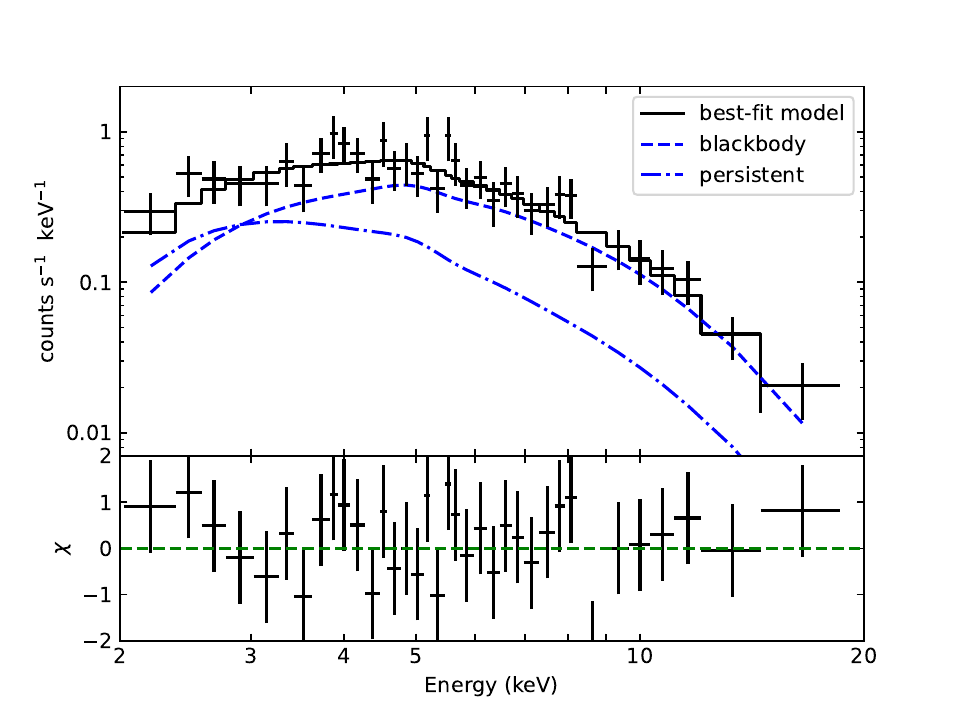}
    \includegraphics[scale=0.54]{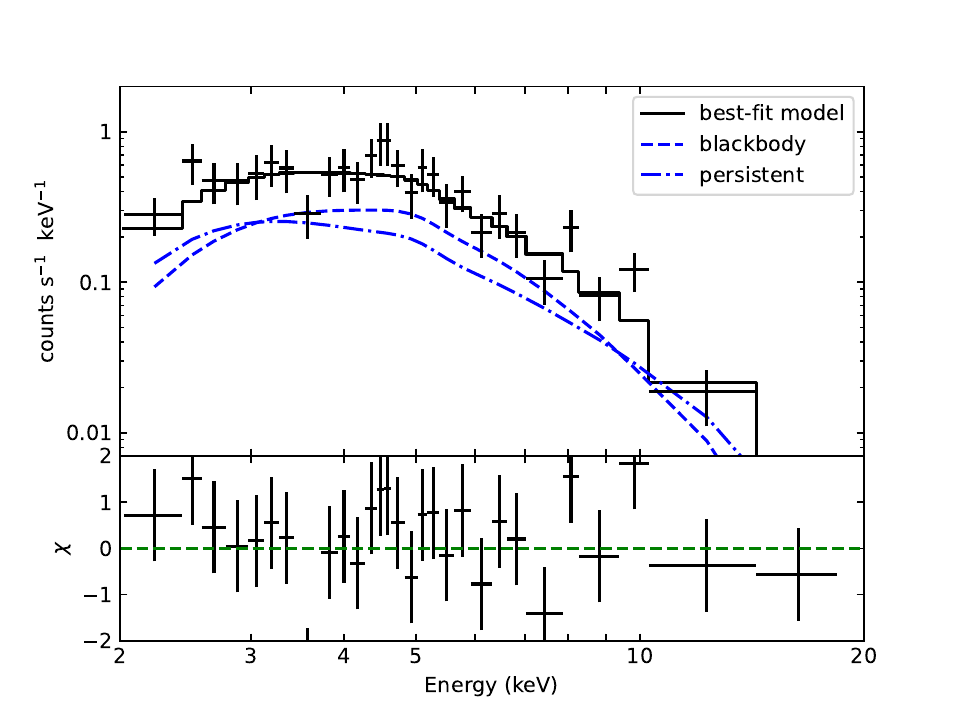}
    \caption{X-ray spectra obtained with MAXI at first (left panel) and second (right panel) scans of  the superburst. Best fit models are also shown with black steps. 
    The dashed and dash-dotted lines in the figures show the BB component and persistent component, respectively.
    Lower panels show the residuals from the best fit model in units of 1-$\sigma$ errors of the data.}
        \label{fig:burstspec}
\end{figure*}

For the spectral analysis, we first fitted the data from 3 scans: 1 scan before, and 2 scans after the burst, using an absorbed the \emph{disk blackbody}~(hereafter~DBB) component and a blackbody~(herafter BB) component. The model parameters are fixed to those of average of the pre-burst emission derived from the analyses in the next section. The scan before the superburst can be adequately modeled without any additional components. However, within the two scans obtained during the burst we see a significant increase in X-ray flux, requiring an additional the \emph{blackbody} component. This additional component indicates the presence of a cooling tail, with temperatures decreasing successively from 1.93$^{+0.23}_{-0.20}$~keV to 1.32$^{+0.20}_{-0.19}$~keV in these orbits, respectively (see \autoref{fig:burstspec}). 

The thermal nature of the additional component and evidence of cooling suggest that this event is indeed the second superburst observed from \src. The resulting total fluxes in 4--10 keV measured before, during and after the burst are 
6.9$^{+0.5}_{-0.8} \times 10^{-9}$ erg s$^{-1}$ cm$^{-2}$ (before burst),
2.5$^{+0.1}_{-0.1} \times 10^{-8}$ erg s$^{-1}$ cm$^{-2}$ (1st scan),
1.4$^{+0.1}_{-0.1} \times 10^{-8}$ erg s$^{-1}$ cm$^{-2}$ (2nd scan),
5.7$^{+0.9}_{-0.7} \times 10^{-9}$ erg s$^{-1}$ cm$^{-2}$ (after burst).
These flux levels, along with the transient thermal component, confirm the superburst nature of the event.

\section{Pointed Observations around the 2020 Superburst}
\label{sec:analaysis}

To investigate the spectral evolution following the 2020 superburst from \src, we analyzed the soft X-ray data obtained with the \nicer X-ray Timing Instrument \citep[XTI,][]{2012SPIE.8443E..13G} and the Insight-HXMT~\citep[HXMT,][]{Li_2007,Zhang_2014,Zhang_2019} observations. We used observations obtained from July 9 to 30, 2020, which covered approximately 15 days of monitoring around the superburst. Unfortunately, neither instrument captured the superburst itself, but both provide valuable pre- and post-superburst data. Both missions observed the source on a daily basis, providing roughly two datasets each day throughout this period. A summary of all observations used in this study is presented in \autoref{tab:sum}. 

\begin{table}
    \centering
    \caption{Summary of all the observations used in this study.}
\begin{tabular}{cccc}
\hline
OBSID & Date$^*$ & Exposure & Instrument \\
 & (MJD) & (ks) & \\
\hline
P020503803801 & 59039.742305 & 1.32 & HXMT \\
P020503803802 & 59039.867757 & 1.02 & HXMT \\
P020503803803 & 59040.007028 & 0.75 & HXMT \\
3657023701 & 59041.003125 & 1.05 & NICER \\
P020503803901 & 59041.266588 & 1.80 & HXMT \\
P020503803902 & 59041.421808 & 1.68 & HXMT \\
P020503803903 & 59041.554273 & 0.78 & HXMT \\
3657023801 & 59042.295787 & 2.64 & NICER \\
P020503804001 & 59042.525986 & 1.80 & HXMT \\
P020503804002 & 59042.680234 & 2.15 & HXMT \\
P020503804003 & 59042.812803 & 0.60 & HXMT \\
3657023901 & 59043.140014 & 0.90 & NICER \\
P020503804101 & 59043.520430 & 2.57 & HXMT \\
P020503804102 & 59043.673891 & 2.03 & HXMT \\
P020503804103 & 59043.806368 & 0.78 & HXMT \\
3657024001 & 59044.350187 & 3.68 & NICER \\
3657024101 & 59044.993451 & 6.84 & NICER \\
P020503804201 & 59044.647710 & 0.08 & HXMT \\
P020503804202 & 59044.799956 & 1.41 & HXMT \\
P020503804203 & 59044.932433 & 0.42 & HXMT \\
3657024002 & 59045.380477 & 1.11 & NICER \\
P020503804301 & 59045.443625 & 2.93 & HXMT \\
P020503804302 & 59045.594828 & 2.18 & HXMT \\
P020503804303 & 59045.727317 & 1.08 & HXMT \\
3657024202 & 59046.992965 & 4.96 & NICER \\
P020503804401 & 59047.235199 & 2.30 & HXMT \\
P020503804402 & 59047.383359 & 2.27 & HXMT \\
P020503804403 & 59047.515835 & 1.32 & HXMT \\
3657024401 & 59048.025639 & 7.41 & NICER \\
P020503804501 & 59048.895014 & 0.72 & HXMT \\
P020503804502 & 59049.039504 & 1.14 & HXMT \\
3657024501 & 59049.517120 & 2.19 & NICER \\
P020503804701 & 59050.422271 & 2.04 & HXMT \\
P020503804702 & 59050.563312 & 1.60 & HXMT \\
P020503804703 & 59050.695835 & 1.00 & HXMT \\
3657024701 & 59051.133660 & 1.91 & NICER \\
P020503804801 & 59051.417768 & 1.82 & HXMT \\
P020503804802 & 59051.578035 & 1.59 & HXMT \\
P020503804803 & 59051.720940 & 0.81 & HXMT \\
3657024801 & 59052.033312 & 6.01 & NICER \\
P020503804901 & 59052.810372 & 0.59 & HXMT \\
P020503804902 & 59052.948995 & 0.90 & HXMT \\
P020503804903 & 59053.081669 & 1.14 & HXMT \\
P020503805001 & 59055.858405 & 2.99 & HXMT \\
P020503805101 & 59057.050812 & 0.76 & HXMT \\
P020503805102 & 59057.203440 & 0.60 & HXMT \\
3657025601 & 59060.332085 & 3.54 & NICER \\

\hline
\end{tabular}
\\
\footnotesize{$^*$ Start time of the observation.}\\
\label{tab:sum}
\end{table}
\subsection{Data Reduction}
\label{sec:data_reduction}

We used \nicer data in the 0.2--10.0 keV energy range and were reduced using the standard filtering \emph{nicerl2} pipeline tool, and spectra were extracted with the \emph{nicerl3-spect} distributed with HEASoft version v6.33.  We grouped our spectra with the optimal binning method~\citep{2016A&A...587A.151K} and tried having at least 100 or 150 counts per channel with similar results. The \emph{nibackgen3C50} model~\citep{2022AJ....163..130R} was used for background estimation.

Because our primary focus is on variations in the soft X-ray band, we specifically concentrated on the LE data from HXMT, and did not use data from the ME and HE detectors. We specifically selected the energy range 2.0--8.0~keV, since the temperature of the LE detector exceeded the valid range for the background model after June 2019, resulting in electronic noise and uncertainties below 2~keV \citep{Li_2020}. 

In HXMT data structure, a single observation, typically lasting several hours, is artificially divided into multiple segments, called "exposures", to reduce data file size. These segments usually last about three hours and are identified by Exposure IDs\footnote{\url{http://hxmten.ihep.ac.cn/SoftDoc/848.jhtml}}. In \autoref{tab:sum}, we provide the detailed exposure times for all such segments used in this work. The data reduction was performed using the Insight-HXMT Data Analysis Software (HXMTDAS) v2.06\footnote{\url{http://hxmtweb.ihep.ac.cn/SoftDoc.jhtml}}, following standard calibration and filtering procedures. 

\subsection{Spectral Modeling}
\label{sec:spectra_modeling}

In a low mass X-ray binary hosting a neutron star usually there are two main regions that contribute to the X-ray emission. One of these components is the inner accretion disk and the other is temperature emitted from the neutron star. Emission from these components is typically Compton up-scattered by a coronal structure, which causes modification in the observed X-ray spectra especially in the hard X-ray part. Additionally to these components reflection features may also be observed due mainly from the coronal emission being reflected by the accretion disk. Using broadband data, \cite{2017MNRAS.467..290A} showed that the X-ray spectra of \src can be modeled with a combination of a disk blackbody and a blackbody components where the second component shows evidence for a Compton up-scattering. In that case they used the {\emph nthcomp} model to take this modification into account, which becomes significant above 10 keV especially when the source is in soft state, similar to the data we analyze here.

In this paper our aim is to detect any impacts of the superburst on the accretion disk; therefore, we mainly focus our attention on the DBB~component, which is dominant in the soft X-rays. Given its relatively broad energy range and much larger effective area towards soft X-rays, we started our fits using only the \nicer data. 

Initially we fit each X-ray spectrum using an absorbed DBB and a BB component. The absorption column density is modeled with a \emph{tbabs} model in {\tt XSPEC} \citep{Arnaud1996} assuming interstellar abundances \citep{2000ApJ...542..914W}. However, we noticed significant edges around 1~keV, which seem to be caused by deviations in the abundance of the interstellar medium from the assumed abundance. Given the larger effective area of NICER, such features can become observable and have been detected in observations of several other bright X-ray sources (see, e.g., \citealt{2020ApJ...895...45L,2023MNRAS.525..595L}). Therefore, we used the {\tt tbvarabs} model and, as a first run, we allowed the Hydrogen column density~($N_{\rm{H}}$) and the abundance of the neon~(Ne) and iron~(Fe) in the ISM to vary. Then we fixed these parameters to their error weighted averages as $N_{\rm{H}} = 1.30\times10^{22}~cm^{-2}$ and the neon abundance and iron abundance as 1.63 and 1.39 times the abundance in the interstellar medium, respectively. We used the same parameters for the absorption column density as fixed parameters when fitting the HXMT LE data. Our final model in {\tt XSPEC} became tbvarabs*(diskbb+bbodyrad).

For the DBB component, following \cite{2017MNRAS.467..290A}, we calculated the inner disk radius using Equation \ref{eq:rin}.

\begin{center}
\begin{minipage}{0.8\linewidth} 
\begin{equation}
  R_{in} = \xi\kappa^{2}\sqrt{\frac{DBB_{Norm}}{\cos{i}}}\times D,
  \label{eq:rin}
\end{equation}
\end{minipage}
\end{center}

where $\xi$ is the correction factor for the inner torque-free boundary condition \citep{1998PASJ...50..667K}, we assumed $\xi=0.4$. $\kappa$ is the color-correction factor and we took as $\kappa=1.7$ \citep{1995ApJ...445..780S}, $i$ is the orbital inclination, which we here assumed to be 70$^{\rm{o}}$, following \cite{2017MNRAS.467..290A}. Finally, $D$ is the distance of the source in units of 10~kpc. We assumed the distance to \src as 4~kpc \citep{2016ApJ...820...28O}. 

We also tested several spectral models on two \nicer X-ray spectra obtained just before and after the superburst (ObsIDs: 3657024002, 3657024202). First of all, we applied a DBB + POW model, however, using such a model resulted in significantly worse $\chi^2$ values. Further, following \cite{2017MNRAS.467..290A}, we also attempted to include the NTHCOMP model~(assuming either DBB or BB as the source of the seed photons) to take into account the above mentioned Compton upscattering. However, an F-test yielded that the apparent improvement to the fit was not statistically significant with a chance probability of 93\%. The likely reason is that, in the soft state, the blackbody emission below 10 keV produces a spectrum very similar to its Compton-upscattered version, and significant differences become apparent only above $\sim 10~keV$, as also seen in Figure 3 of \cite{2017MNRAS.467..290A}.

\subsection{Results and The Evolution of Spectral Parameters}
\label{sec:spectra_evolution}

\begin{figure*}
     \centering 
    \includegraphics[scale=0.32]{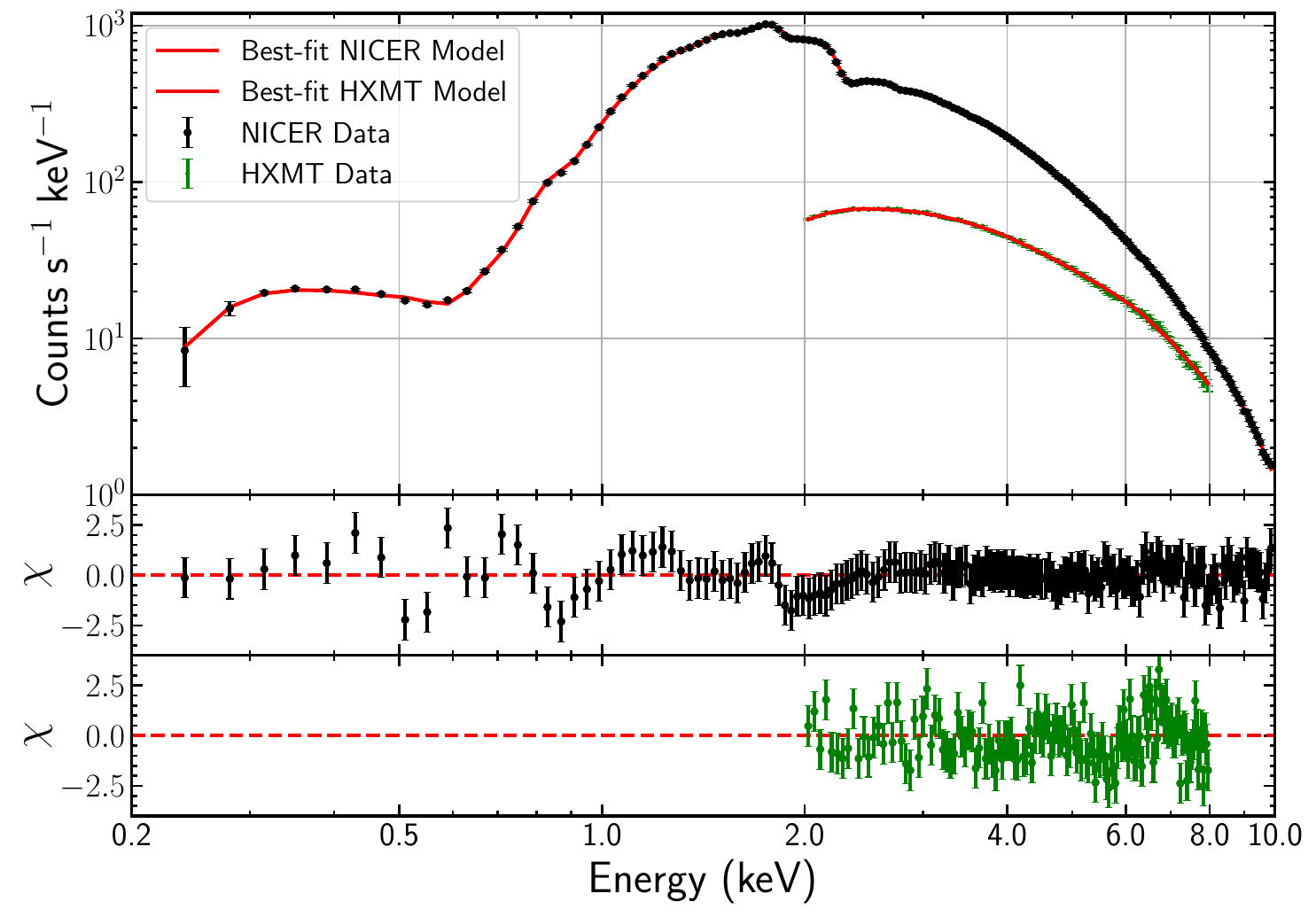}
    \includegraphics[scale=0.32]{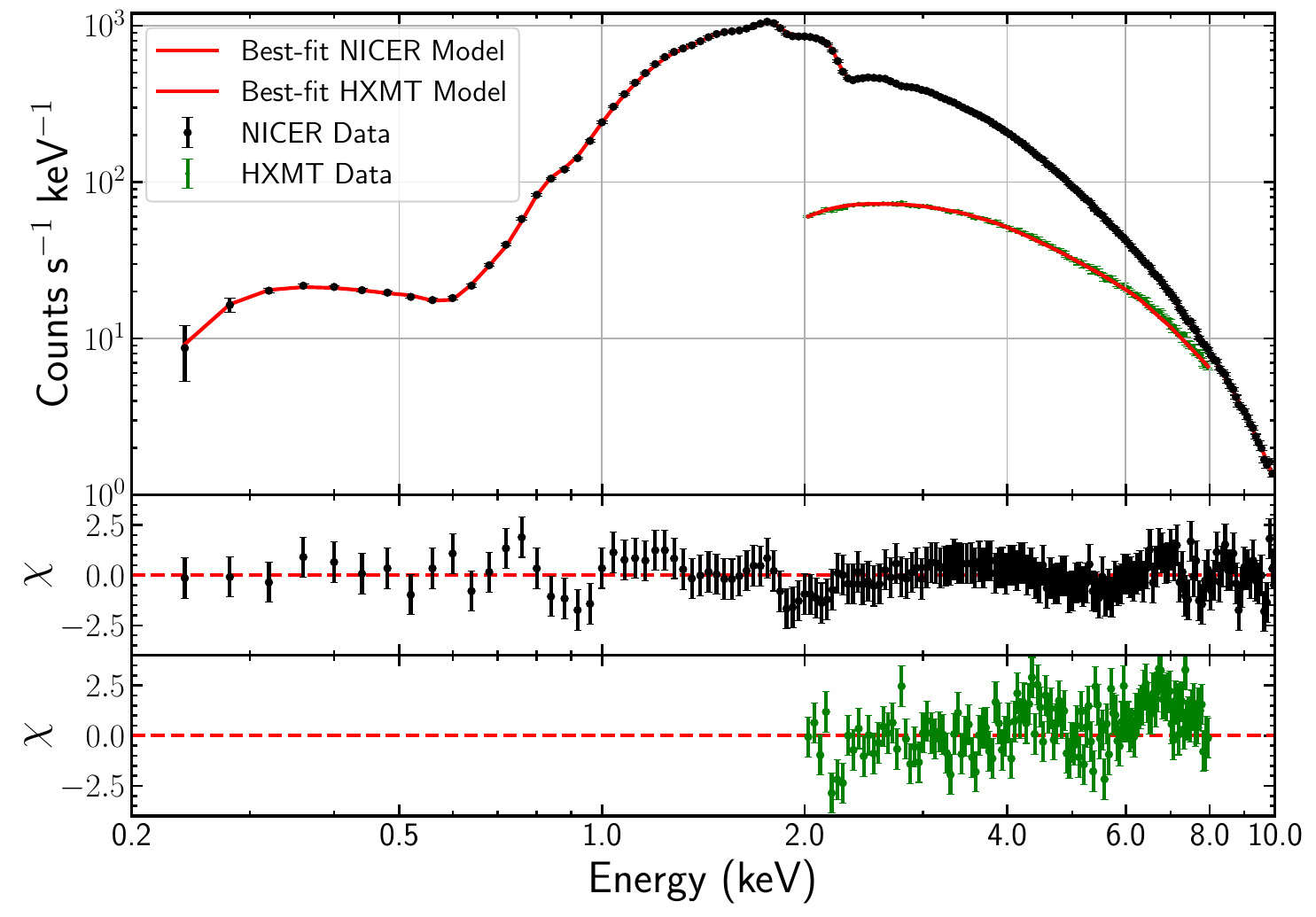}
    \caption{Example X-ray spectra obtained with \nicer and HXMT two days before~(left panel) and just after~(right panel) the superburst. Best fit models are also shown with thick red lines. Lower panels show the residuals from the best fit model in units of 1-$\sigma$ errors of the data. For plotting purposes, HXMT data is rebinned additionally with a factor of 5.}
        \label{fig:spectrum}
\end{figure*}

\begin{figure*}
     \centering 
    \includegraphics[scale=0.40]{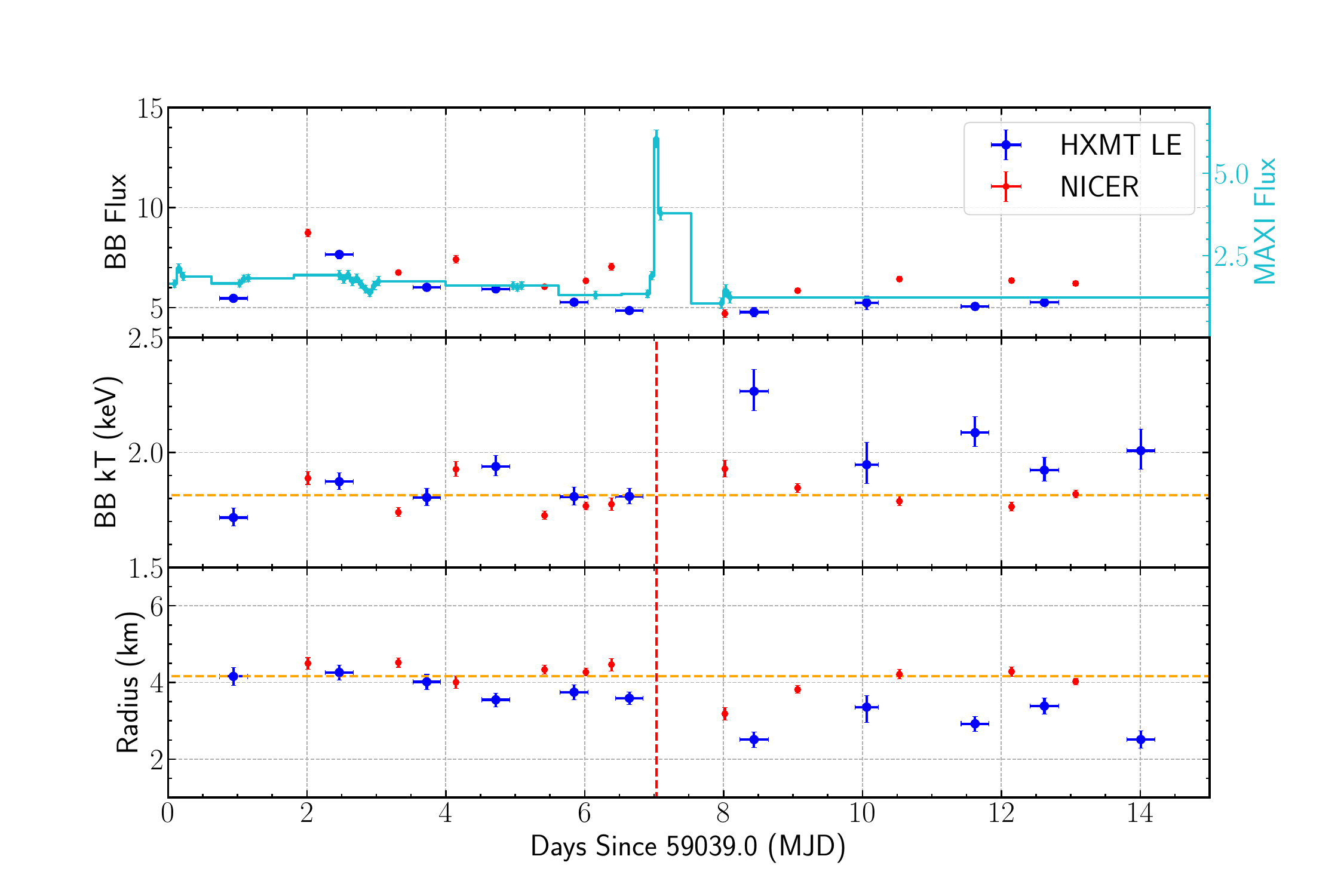}
    \includegraphics[scale=0.40]{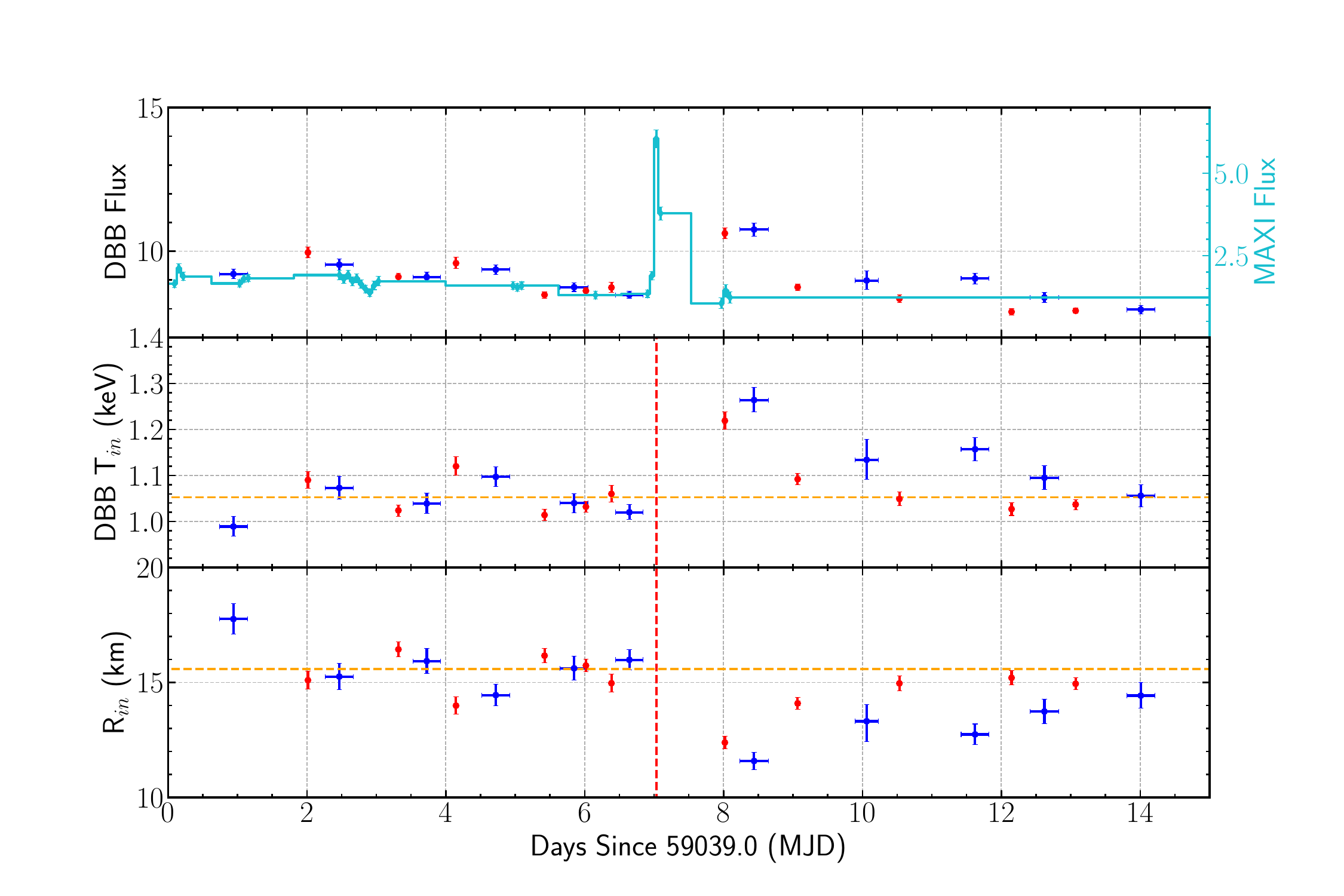}
    \caption{The spectral evolution of \src is shown for the BB component (upper panel) and the DBB component (lower panel) as inferred by \nicer (red) and Insight-HXMT (blue). Upper panels show the inferred unabsorbed flux (0.2--10.0~keV for \nicer and 2--8~keV for HXMT) of the BB and DBB components in units of $\times10^{-9}$\fluxcgs as well as the MAXI fluxes in 2-20~keV~ph~s$^{-1}$~cm$^{-2}$ (right y axis in cyan). Middle panels show the blackbody and the inner disk temperatures. Finally, lower panels show the apparent emitting radius and the apparent inner disk radius in units km. The inner disk radius is calculated using \autoref{eq:rin}. In all panels yellow horizontal dashed lines show the average values obtained before the superburst. The red vertical lines show the date of the super burst as inferred from the MAXI data.}
        \label{fig:spec_ev}
\end{figure*}

In \autoref{fig:spectrum}, we show two examples of \nicer and HXMT X-ray spectra with their best-fit models, taken two days before and just after the superburst. The data points shown in \autoref{fig:spectrum} have been rebinned for clarity. In \autoref{fig:spec_ev} and \autoref{tab:fit_res}, we show the time evolution of the spectral parameters and the resulting best-fit parameters of the models, respectively. For the spectral fits of the HXMT data, we fit the data from individual segments simultaneously by linking all the parameters. In most cases, this approach resulted in better constrained spectral parameters with acceptable fits (with an average $\chi^2_{\nu}=1.23$). Because of this consideration, the errors in the x-axes of Figures \ref{fig:spec_ev} and \ref{fig:decay_mod} reflect not just the exposure times, but rather the total duration of the HXMT observations including the gaps in between the segments. The spectral evolution can be investigated through changes in the parameters of the DBB component, particularly its temperature and normalization. The average inner disk radius before the superburst can be calculated as 15.5 km and the inner disk temperature are 1.04~keV, which is similar to what has been reported by \cite{2017MNRAS.467..290A}. Overall fluxes of both the DBB and the BB components show a slow decline as expected from the decline of the outburst (see \autoref{maxi_lc}). However, a systematic and significant variation in flux, temperature, and inner disk radius parameters of the DBB component can be seen just after superburst. A similar but much smaller change in the temperature and emitting radius of the BB component can also be seen. 

\begin{table*}
    \centering
    \caption{Best fitting model results of \src using an absorbed DBB and a BB model. Red horizontal line indicates the rough start time of the superburst. Note that in order to be able to obtain more constrained parameters, we performed a simultaneous fit with the HXMT data and linked all the parameters of each segment.}
    \begin{tabular}{cccccccc}
\hline
Date$^*$ & T$_{in}$ & R$_{in}$ & DBB Flux$\dagger$ & kT & R$_{app}$ & BB Flux$^\dagger$ & $\chi^2$ / dof \\
(MJD)  & (keV)    & (km) & & (keV) & (km) & & \\
\hline
 59039.942030 & $0.99\pm0.02$ & $17.75^{+0.67}_{-0.64}$ & $9.21^{+0.17}_{-0.16}$ & $1.72\pm0.04$ & $4.16\pm0.23$ & $5.46\pm0.19$ & 2650.33/2120 \\
59041.015285 & $1.09\pm0.02$ & $15.10^{+0.39}_{-0.38}$ & $9.96^{+0.20}_{-0.19}$ & $1.89\pm0.03$ & $4.50\pm0.15$ & $8.73\pm0.19$ & 99.69/160 \\
59041.467000 & $1.07^{+0.03}_{-0.02}$ & $15.26^{+0.57}_{-0.56}$ & $9.53^{+0.20}_{-0.19}$ & $1.87\pm0.04$ & $4.26^{+0.19}_{-0.20}$ & $7.65^{+0.20}_{-0.22}$ & 3021.67/2120 \\
59042.318165 & $1.02\pm0.01$ & $16.44^{+0.32}_{-0.31}$ & $9.11^{+0.13}_{-012}$ & $1.74\pm0.02$ & $4.52\pm0.12$ & $6.75\pm0.13$ & 108.32/164 \\
59042.726275 & $1.04\pm0.02$ & $15.93^{+0.55}_{-0.54}$ & $9.10^{+0.17}_{-0.16}$ & $1.80\pm0.04$ & $4.02^{+0.19}_{-0.20}$ & $6.01^{+0.18}_{-0.19}$ & 2889.61/2120 \\
59043.147590 & $1.12\pm0.02$ & $13.99^{+0.38}_{-0.37}$ & $9.59^{+0.20}_{-0.19}$ & $1.93^{+0.04}_{-0.03}$ & $4.00\pm0.16$ & $7.41^{+0.19}_{-0.20}$ & 106.20/156 \\
59043.720765 & $1.10\pm0.02$ & $14.45^{+0.47}_{-0.46}$ & $9.36^{+0.17}_{-0.16}$ & $1.94^{+0.05}_{-0.04}$ & $3.55\pm0.18$ & $5.93^{+0.18}_{-0.19}$ &3174.46/2120\\
59044.422640 & $1.01\pm0.01$ & $16.16\pm0.30$ & $8.48^{+0.12}_{-0.11}$ & $1.73\pm0.02$ & $4.34\pm0.12$ & $6.04\pm0.11$ & 100.46/164 \\
59045.018505 & $1.04\pm0.02$ & $15.61^{+0.52}_{-0.51}$ & $8.75^{+0.16}_{-0.15}$ & $1.81\pm0.04$ & $3.75\pm0.19$ & $5.26^{+0.16}_{-0.17}$ & 2132.53/2109 \\
59044.848095 & $1.03\pm0.01$ & $15.74\pm0.27$ & $8.62^{+0.11}_{-0.0}$ & $1.77\pm0.02$ & $4.27\pm0.10$ & $6.34^{+0.10}_{-0.11}$ & 91.53/170 \\
59045.386070 & $1.06\pm0.02$ & $14.97\pm0.39$ & $8.74^{+0.18}_{-0.17}$ & $1.78\pm0.03$ & $4.47\pm0.16$ & $7.04^{+0.17}_{-0.18}$ & 96.01/158 \\
59045.643885 & $1.02\pm0.02$ & $15.98^{+0.45}_{-0.43}$ & $8.48^{+0.12}_{-0.11}$ & $1.81^{+0.04}_{-0.03}$ & $3.59\pm0.16$ & $4.85\pm0.13$ &3352.74/2120 \\
\arrayrulecolor{red}\hline
\arrayrulecolor{black}
59047.019300 & $1.22\pm0.02$ & $12.39^{+0.28}_{-0.27}$ & $10.62\pm0.02$ & $1.93\pm0.04$ & $3.18\pm0.16$ & $4.69^{+0.18}_{-0.19}$ & 98.21/168 \\
59047.439945 & $1.26\pm0.03$ & $11.59^{+0.38}_{-0.37}$ & $10.85\pm0.02$ & $2.27^{+0.10}_{-0.08}$ & $2.51\pm0.20$ & $4.77^{+0.22}_{-0.23}$ &2334.41/2120 \\
59048.067575 & $1.09\pm0.01$ & $14.09\pm0.25$ & $8.75^{+0.12}_{-0.11}$ & $1.85\pm0.02$ & $3.81\pm0.10$ & $5.84\pm0.11$ & 89.11/171 \\
59049.061330 & $1.13^{+0.05}_{-0.04}$ & $13.31^{+0.74}_{-0.88}$ & $8.98^{+0.34}_{-0.32}$ & $1.95^{+0.10}_{-0.08}$ & $3.36^{+0.30}_{-0.40}$ & $5.24^{+0.33}_{-0.36}$ &1540.73/1412\\
59049.533220 & $1.05^{+0.02}_{-0.01}$ & $14.96^{+0.32}_{-0.31}$ & $8.35\pm0.13$ & $1.79\pm0.02$ & $4.21\pm0.12$ & $6.42\pm0.13$ &91.98/162\\
59050.622840 & $1.16\pm0.03$ & $12.74^{+0.45}_{-0.44}$ & $9.05^{+0.19}_{-0.18}$ & $2.08^{+0.07}_{-0.06}$ & $2.92\pm0.19$ & $5.05^{+0.19}_{-0.20}$ & 2566.86/2120 \\
59051.148075 & $1.03\pm0.01$ & $15.20\pm0.32$ & $7.89\pm0.12$ & $1.76\pm0.02$ & $4.28\pm0.12$ & $6.35^{+1.18}_{-1.22}$ &136.15/161 \\
59051.620950 & $1.10\pm0.03$ & $13.74^{+0.54}_{-0.52}$ & $8.39^{+0.19}_{-0.18}$ & $1.92^{+0.06}_{-0.05}$ & $3.38^{+0.20}_{-0.21}$ & $5.25^{+0.19}_{-0.20}$ &2264.40/2120\\
59052.071260& $1.04\pm0.01$ & $14.94\pm0.25$ & $7.93\pm0.09$ & $1.82\pm0.02$ & $4.03\pm0.09$ & $6.20\pm0.09$ &105.07/169\\
59053.009520 & $1.06\pm0.02$ & $14.43^{+0.57}_{-0.54}$ & $7.97\pm0.15$ & $2.01^{+0.09}_{-0.08}$ & $2.51\pm0.23$ & $3.31\pm0.16$ &2412.14/2118 \\
59056.058560 & $1.01\pm0.02$ & $15.44^{+0.62}_{-0.59}$ & $7.52\pm0.14$ & $1.83^{+0.06}_{-0.05}$ & $3.17\pm0.22$ & $3.96^{+0.15}_{-0.16}$ &673.84/704\\
59057.250955 & $1.02\pm0.03$ & $15.16^{+0.82}_{-0.75}$ & $7.49\pm0.17$ & $1.93^{+0.11}_{-0.09}$ & $2.63^{+0.30}_{-0.29}$ & $3.18\pm0.20$ &1471.89/1400\\
59060.355135 & $1.04\pm0.03$ & $14.26^{+0.62}_{-0.59}$ & $7.31\pm0.15$ & $1.89^{+0.07}_{-0.06}$ & $2.92\pm0.22$ & $3.72^{+0.16}_{-0.17}$ &26.28/98 \\
\hline
\end{tabular}\\ 
\footnotesize{$^*$ Midpoint of the combined time span of the all the observation segments for Insight-HXMT.}\\
\footnotesize{$^\dagger$ Unabsorbed 0.5$-$10~keV flux in units of $\times$10$^{-9}$~erg s$^{-1}$ cm$^{-2}$.}\\
    \label{tab:fit_res}
\end{table*}

\section{Discussion}
\label{sec:discussion}
\subsection{Comparison of the 2020 Outburst with Previous Outbursts}
\label{sec:comparison1}

\begin{figure}
    \centering
    \includegraphics[scale=0.55]{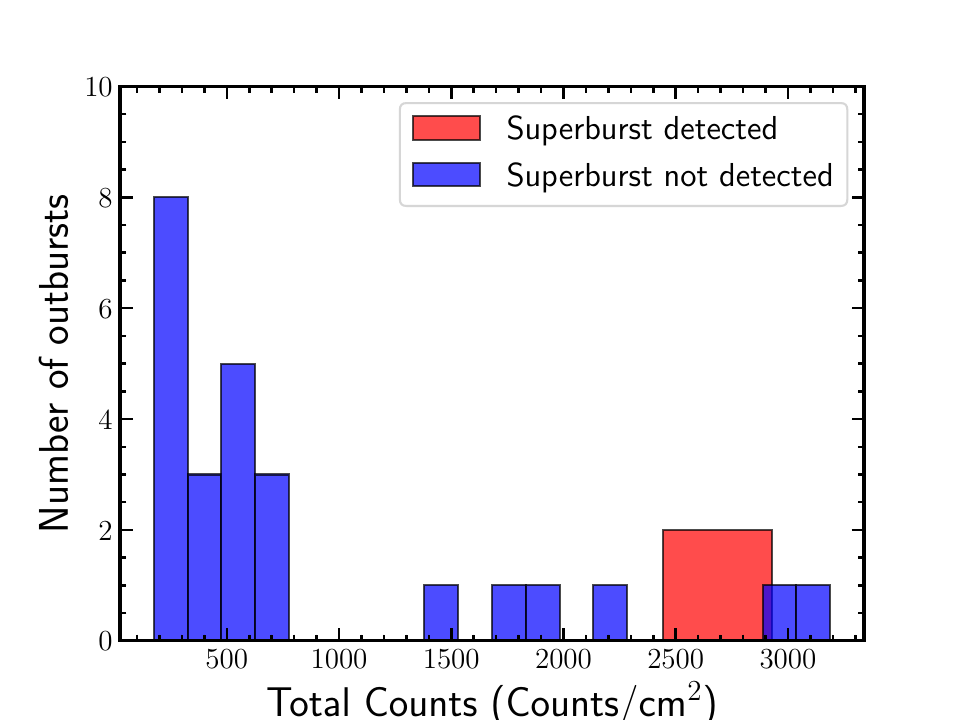}
    \caption{Fluences of all the outbursts detected from \src by RXTE/ASM or MAXI as calculated from observed count rates.}
    \label{fig:fluence}
\end{figure}

\begin{figure*}
\centering
\includegraphics[scale=0.8]{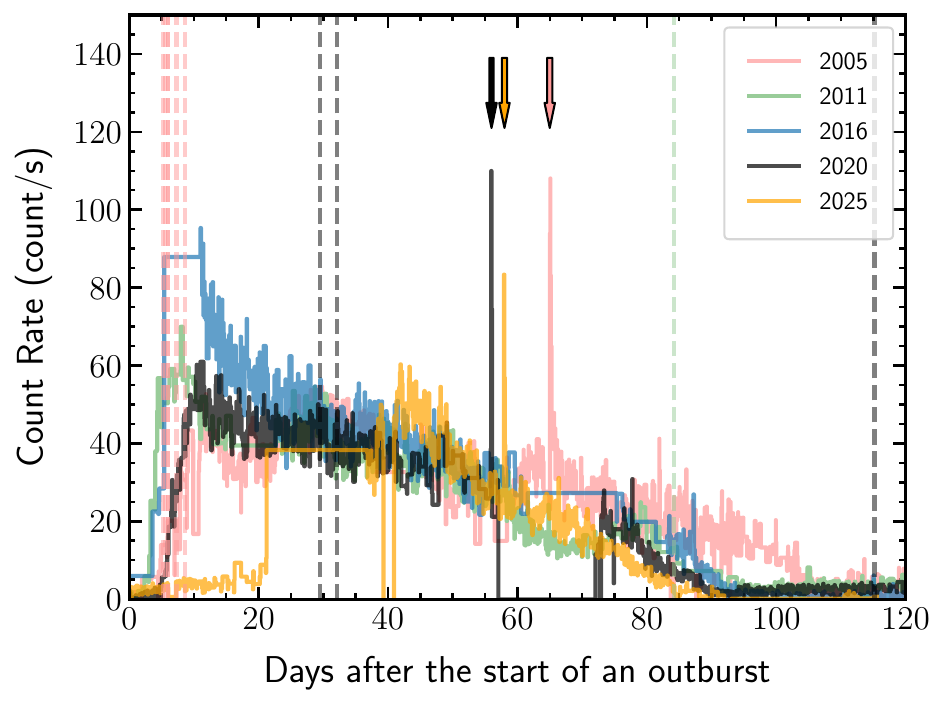} 
\caption{2--10~keV lightcurves of the five outbursts observed from \src~with ASM (2005) and MAXI (2011, 2016 and 2020) showing similar profiles. Since MAXI data is given in units of ph~s$^{-1}$~cm$^{-2}$, we multiplied MAXI values with a factor of 22 \protect\citep{2014MNRAS.439.2717G} to match the ASM count rates. Superbursts in 2005, 2020, and 2025 outbursts are shown via vertical arrows. Note that the start of the 2025 outburst is not observed, to be able to compare with other superbursts we moved the 2025 data by about 40 days. Type-I burst times for 2005 and 2011 outbursts are also marked with vertical dashed lines (with matching colors) as reported in the MINBAR catalog \protect\citep{2020ApJS..249...32G}. While no bursts during the 2016 outburst from \src is reported, for 2020 outburst we used the times given in \protect\cite{2021ApJ...910...37G}. }
\label{fig:outburst_comp}
\end{figure*}

\src~shows frequent outbursts with a typical recurrence time of 330 days \citep{simon_2004}. To understand the conditions that led to a superburst in \src it is important to compare the 2020 outburst with previous outbursts from the source.  Despite relatively frequent outbursts, the 2020 outburst was one of the longest and brightest \citep{simon_2004, 2024arXiv240718867H}. In particular, the 2020 outburst was very much similar to the outburst observed in May 2005 during which the first superburst from \src~was detected \citep{Keek_2008}. In order to see if the outbursts of \src from, where a superburst has been observed, are different from others we calculated roughly the total fluences of 27 outbursts reported by \cite{2024arXiv240718867H}, observed since 1998. For that purpose, we obtained daily averaged data from RXTE/ASM and MAXI (multiplied by a factor of 22 to roughly match the ASM count rate, \cite{2014MNRAS.439.2717G}) and smoothed the lightcurves by means of a spline interpolation method. We then calculated the integral of the resulting function. Note that for 2005 and 2020 we did not use the data points covering the superbursts. We applied an interpolation to recover visibility gaps in all outbursts. A histogram of the total counts detected from all the outbursts is given in \autoref{fig:fluence}. Out of the 27 outbursts the total integral yields the largest total counts from outbursts that occurred in 2005, 2011, 2016, and 2020 with 2921, 2665, 3188, and 2579 counts, respectively. In \autoref{fig:outburst_comp}, we show these outbursts using data from RXTE/ASM and MAXI. Since in an accreting system the total luminosity can be attributed to the mass accretion rate, the fluence during an outburst can also be attributed to roughly the total mass accreted onto the neutron star. In that manner, we see that the total fluence and therefore the amount of matter accreted is largest in these four outbursts.

The first superburst observed from \src~was detected on the 55th day of the outburst in 2005 \citep{Keek_2008}, whereas the superburst in 2020 happened on the 51st day, suggesting a potential link between the accumulation of accreted material and the ignition conditions of a superburst. We calculated that 73 and 82 \% of the total counts were detected before the superbursts in 2005 and 2020, respectively. For reference, at these times the fluence of the outbursts reached to 2132.33 and 2114.78 counts, which is still much larger than most of the outbursts shown in \autoref{fig:fluence}. 
While Type-I bursts do not affect significantly the total amount of accreted matter, they may influence the composition of the accreted layer by producing carbon-rich ashes. This process can be relevant for accumulating enough carbon to fuel a superburst \citep{Cumming_2001,Strohmayer_2002}.
While during the rise of the 2005 outburst, 6 Type-I X-ray bursts were observed, only 2 bursts were observed in 2020 before the superburst. We note that, in 2020 despite a very good coverage of the entire outburst, the initial rise phase is not covered by \nicer \citep[see Figure 1 of][]{2021ApJ...910...37G}. 

\begin{figure}
     \centering 
    \includegraphics[scale=0.50]{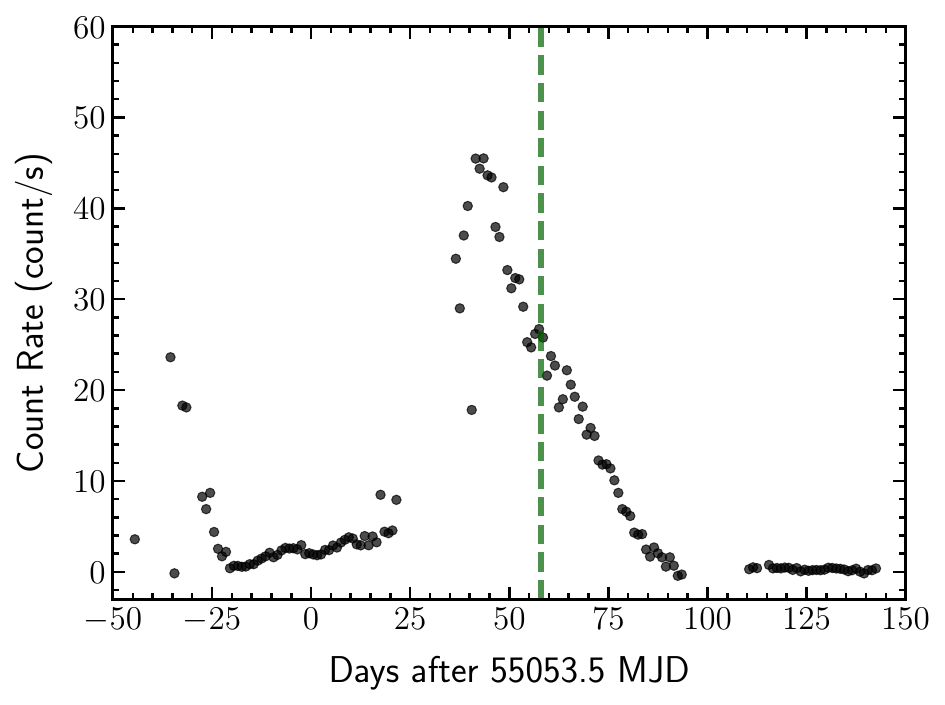}
    \caption{2--10 keV lightcurve of the 2025 outburst of \src as observed with MAXI. Similar to other figures the MAXI flux values were scaled by a factor of 22 \protect\citep{2014MNRAS.439.2717G} to match the approximate ASM count rate levels. A vertical dashed green line indicates the time of the superburst observed during this outburst.}
        \label{fig:outburst_only2025}
\end{figure}

In 2025, a new superburst was detected by MAXI \citep{Serino_2025}. The outburst during which this superburst occurred was completely different from previous outbursts (see~\autoref{fig:outburst_only2025}). Before showing a more traditional outburst profile, in around mid-December 2024, \src showed a mini outburst, of which the beginning could not be observed, but the decay was detected to last at least 10 days. Starting with the new year, MAXI detected a gradual increase in its X-ray brightness, which lasted approximately forty days. Due to limited coverage, it is hard to determine the exact date, but sometime in late February 2025, \src showed a more classical fast rising outburst, which decayed by the end of April 2025. For comparison, if we perform the above calculation taking into account only the last part, up to April 2025, the total fluence yields 1301 counts from \src, which is significantly lower than the 2005, 2011, 2016, and 2020 outbursts, we show this part of the outburst in \autoref{fig:outburst_comp}. However, if we start the integration from December, including the mini outburst and the continuous flux increase, the fluence reaches to 2051 counts, which is similar to the values reached in 2005 and 2020 outbursts before the superbursts occurred. We should note that although the amount of accreted total mass before a superburst is triggered can be an important factor, there are other factors such as the immediate accretion rate and chemical composition for example the accumulated carbon amount, which may prepare the conditions that would trigger a superburst \citep[see e.g.,][]{2004NuPhS.132..435C, 2006csxs.book..113S}.

\subsection{Superbursts from \src}
\label{sec:comparison}

The detection of a significant X-ray flux increase on the 51st day of the 2020 outburst provides strong evidence for a superburst in \src. MAXI observations reveal a sudden flux increase by a factor of 4.5, followed by a gradual decay over the next few hours. The rapid increase in flux, combined with the emergence of an additional thermal component in the spectral analysis, strongly suggests that the observed event was a superburst. The subsequent cooling phase, inferred from time-resolved spectral analysis, further supports this classification. These characteristics are consistent with previously detected superbursts in other sources, such as 4U~1820--30 \citep{Strohmayer_2002} and KS~1731--260 \citep{Kuulkers_2002}, where similar temporal profiles and spectral evolutions have been reported.

During another outburst in 2025, \src showed a new superburst. The light curve of this superburst shows the characteristic fast-rise exponential-decay profile \citep{Serino_2025}. In addition, the spectral analysis of the MAXI scans after the event shows significant cooling from 1.7~keV to 1.2~keV within 3 hours. The combination of the light curve shape and the detected cooling indicate the presence of a new superburst \citep{Serino_2025} from \src.

\begin{figure}
    \centering
    \includegraphics[scale=0.55]{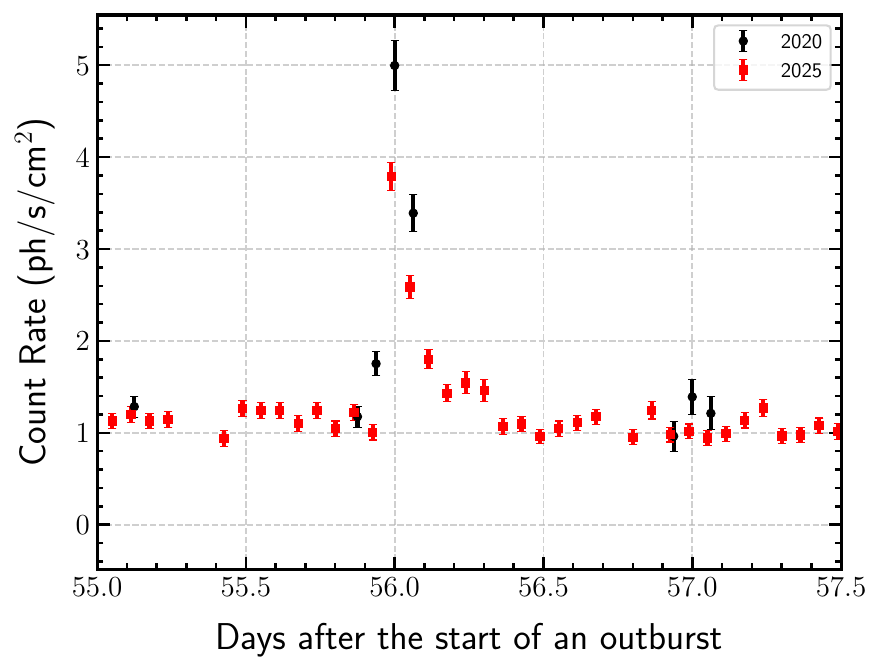}
    \caption{2--10~keV MAXI orbital light curves of 2020 and 2025 superbursts observed from \src. It can be seen that 2020 outburst peaks at a higher rate despite the much sparse coverage especially during cooling.}
    \label{fig:2020_2025superburst}
\end{figure}

In \autoref{fig:2020_2025superburst} we show the MAXI light curves of both superbursts matching their start time. In the 2--20~keV band, the measured peak flux of 2020 superburst is 6.05 photons/s/cm$^{2}$ whereas the peak reached in 2025 is 4.78 photons/s/cm$^{2}$, showing that the 2020 superburst was brighter at its peak by at least 26\%. We note that in both cases the exact peak moments of the superbursts may be missed by MAXI given its 90 minute cadence. However, the measurements present at least a lower limit on the peak fluxes.

\begin{figure}
    \centering
    \includegraphics[scale=0.55]{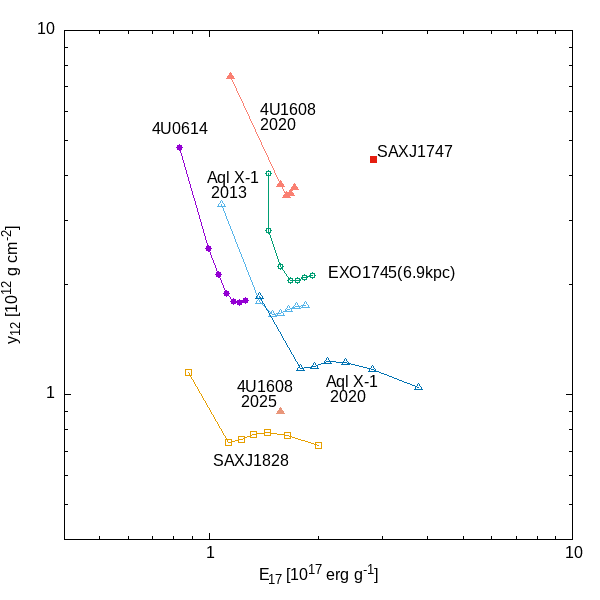}
    \caption{Ignition depth~(y$_{12}$) and released energy~(E$_{17}$) of superbursts from several sources calculated following using Equation~(4) of \citep{Cumming_2004}. Values for different sources are obtained from \citep{2016PASJ...68...95S}.}
    \label{fig:ignition_depth}
\end{figure}

Assuming a distance of 4~kpc for \src~\citep{2016ApJ...820...28O}, we also estimated the released energy per unit mass (E$_{17}$, in units of~$10^{17}$ erg~g$^{-1}$) and the ignition column depth (y$_{12}$, in units of~$10^{12}$~g cm$^{-2}$) using Equation~(4) of \cite{Cumming_2004}. \autoref{fig:ignition_depth} shows the ignition depth and the released energy, including the superbursts observed in 2020 and 2025, in comparison with some of the superbursts observed from different sources from \cite{2016PASJ...68...95S}. Given the small number of observations of the 2020 superburst these calculations present only a lower limit. Still the mass fraction of burnt carbon of 10-20 \%, is very typical compared to other sources \citep{2017symm.conf..121I}. However, note that the ignition depth calculated for the 2020 superburst is one of the largest with respect to what is inferred for other sources in \cite{2016PASJ...68...95S} or in \cite{2017symm.conf..121I}. For the 2005 superburst \cite{Keek_2008} calculated an ignition column depth of y$_{12}$ = 1.5–4.1~g cm$^{-2}$ and an energy release per unit mass of E$_{17}$ = 1.5–1.9. These show that while the E$_{17}$ values are typically the same in all three superbursts the ignition depth varies significantly. The agreement in the released energy in different superbursts supports the idea that the chemical composition of the burning material is similar, however the ignition depth varies as a function of the waiting time before a superburst.

It is known that superbursts result in a temporary cessation of normal type-I X-ray bursts, known as quenching. This is caused by the envelope being too hot for unstable nuclear burning, leading to delay in the resumption of type-I bursts \citep{Kuulkers_2002,Keek_2011, 7600607c}. In this case, a normal type-I X-ray burst is detected from \src on September 13th \citep[see Figure 1 of ][]{guver2021thermonuclear-868}, so the quenching time can be estimated as 58.98 days, which is much shorter than the previous upper limit (99.8 days) reported in 2005 \citep{Keek_2008}, but longer than the 9.44 days reported by \cite{Li_2020} for Aql X-1.

It is known that superbursts result in a depletion of the accreted material necessary for normal bursts to occur for some time. In this case, a normal type-I X-ray burst is detected from \src~on September 13th (see \autoref{fig:outburst_comp}; \citealt{guver2021thermonuclear-868}), so the quenching time can be estimated as 58.98 days, which is much shorter than the previous upper limit (99.8 day) in 2005 \citep{Keek_2008}, but longer than the 9.44 days reported by \cite{2021ApJ...920...35L} for Aql~X-1. The variation in the burst quenching times indicates that the post-burst environment and cooling processes may differ depending on variations in accretion history, local fuel composition, and the thermal state of the neutron star crust \citep{2016PASJ...68...95S,2020ASSL..461..209G}. However note that reported burst quenching times are always upper limits since there is not continuous monitoring of these sources. For the 2025 superburst no such burst data is available yet. 

\subsection{Accretion Disk Evolution Around the Time of the 2020 Superburst}
\label{sec:accretion}

Due to their immense and prolonged radiation, superbursts are thought to impose strong impact on the surrounding accretion environment. To probe such phenomena, we examined the spectral evolution of the accretion disk after the 2020 superburst. 

\begin{figure*}
    \centering
    \includegraphics[scale=0.5]{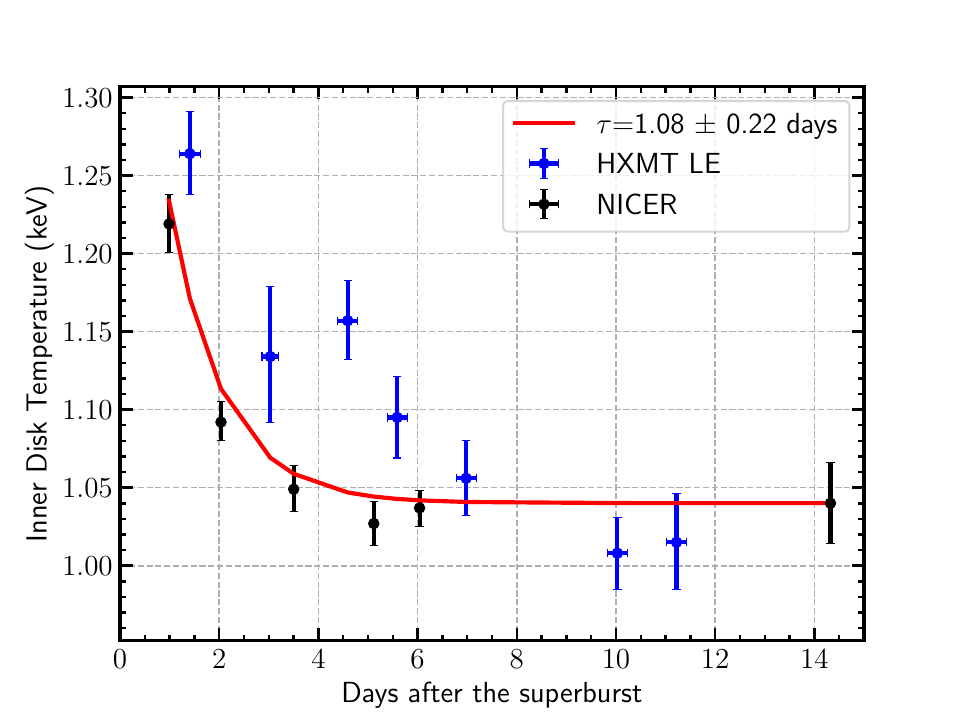}
    \includegraphics[scale=0.5]{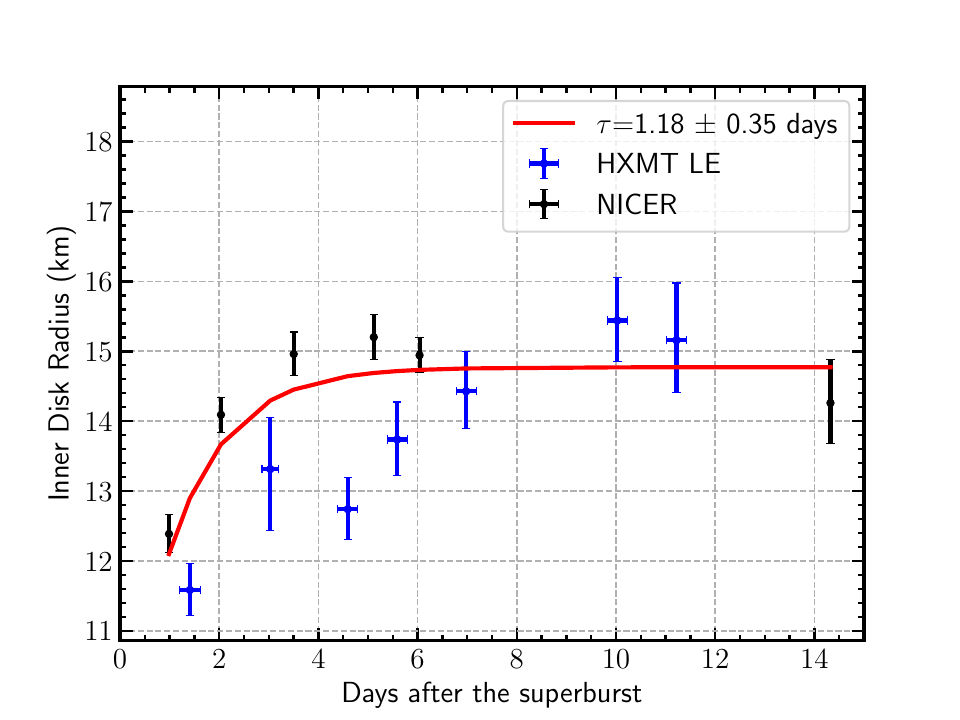}
    \caption{Inner disk temperature and radius values obtained after the superburst time. In order to characterize the spectral evolution in the DBB model we fit the inner disk temperature and radius parameters with exponential cooling, and recovery models, respectively. In both cases we find that within approximately three days the parameters returned to their average value.}
    \label{fig:decay_mod}
\end{figure*}

Although \nicer performed 78 observations of the 2020 outburst with a total exposure time 278~ks \citep{guver2021thermonuclear-868} unfortunately, it could not observe the superburst itself. Assuming that the superburst occured at MJD 59046 (16th July 2020), the closest observation was performed roughly 23 hours and 32 hours after with \nicer and HXMT, respectively. Still, these data show evidence of spectral variation especially in the form of a cooling disk blackbody component~(see Figures~\ref{fig:spec_ev} and \ref{fig:decay_mod}). Our spectral analysis reveals that before the superburst, the inferred inner disk temperature was around 1.04 keV. In the first dataset obtained after the superburst we measure an inner disk temperature of ~1.25 keV, which cools down to the average in a few days. Similarly, the inner disk radius showed a significant drop from 15.5~km to 11.59~km. The observed inner disk temperature and radius variations indicate that the superburst may have affected the inner accretion flow, supporting the idea that superbursts affect the surrounding accretion disk, leading to transient modifications in its geometry and temperature \citep{Keek_2011,2018MNRAS.480....2T}. We would like to note that the parameters of the disk blackbody and the blackbody models we use to fit the data may be somewhat correlated with each other. To better see if it the disk blackbody component is varying, took the first NICER dataset obtained after the superburst (Obsid: 3657024202) and fitted it with the DBB parameters from the previous observation (pre-superburst, Obsid:3657024002). We saw that the post-superburst spectrum can not be fitted by only allowing the BB component to vary, with a resulting $\chi^2$=303.42 for 170 dof as opposed to the $\chi^2$=98.21 for 168 dof.

In \autoref{fig:decay_mod}, we show the evolution of the inner disk radius and temperature parameters after the superburst. Although there seems to be a systematic difference in between the values inferred from HXMT LE and NICER, in both cases a cooling and a recovery in the inner disk radius is observed.  We simultaneously fit the data from both missions with an exponential function to determine the cooling/recovery time scales for the inner disk temperature and radius, respectively. In both cases, we obtained an e-folding time of about 1 day. Note that using only HXMT data results in a slower return to average values. The systematic difference between the two instruments can be attributed to differences in the effective areas of the two instruments. Although the HXMT LE is mostly sensitive to the 2--8~keV energy band, NICER probes a broader range covering 0.2--10~keV, with a significantly larger effective area. However it is worth noting that there seems to be a time dependent variation in the observed differences. For example, while the disk blackbody parameters seem to agree with each other before the superburst (with a average difference of about 0.7 $\sigma$ between the two satellites) the difference increases after the superburst to about 1.9 $\sigma$, which are the parameter values shown in \autoref{fig:decay_mod}. The differences become more evident in the inferred parameters of the blackbody component. Such a time dependence may be suggesting short term (within about a day) hard X-ray variability that our analysis can not probe in detail given the times between the observations of HXMT and NICER.

\begin{figure*}
    \centering 
    \includegraphics[scale=0.45]{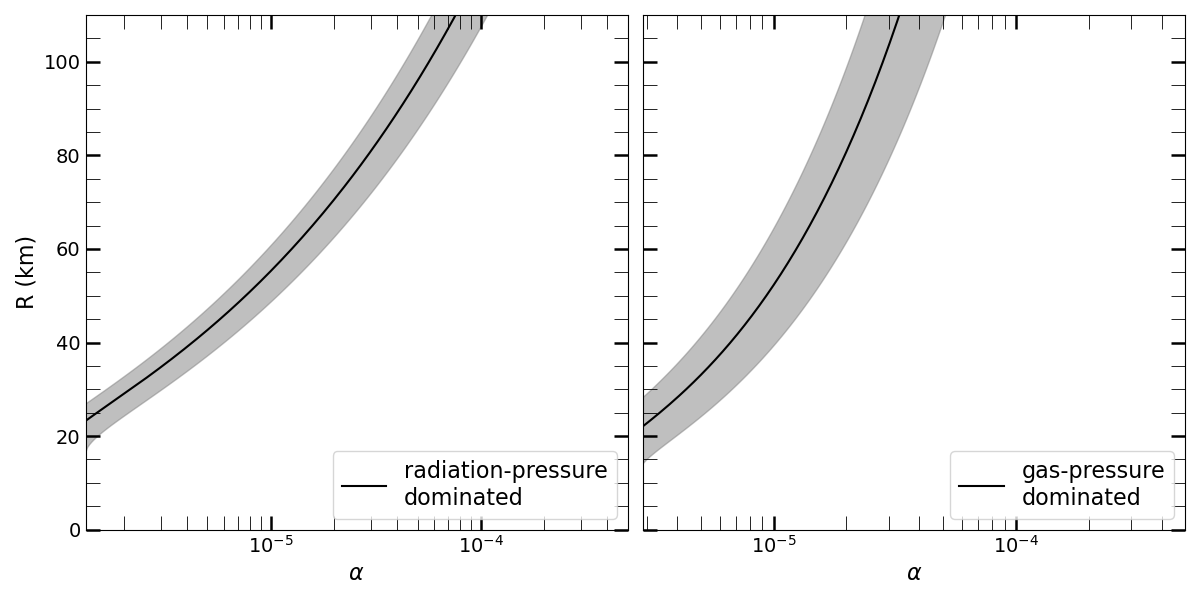}
    \caption{Radius versus viscosity parameter $\alpha$ for a radiation-pressure dominated (left panel, \autoref{eq:nu_rad}) and a gas-pressure dominated disk (right panel, \autoref{eq:nu_gas}). The black lines give the radius for a given $\alpha$ and a viscous timescale of $\tau=1.18$ days, the grey shaded areas were calculated considering a $0.35$ day uncertainty in $\tau$ (see right panel of \autoref{fig:decay_mod}). The change in inner disk radius displayed in the right panel of \autoref{fig:decay_mod} due to viscous processes would require $\alpha\lesssim 10^{-6}$.}
        \label{fig:rad_disk_sim}
\end{figure*}

Variations in disk temperature and flux suggest that irradiation of the disk due to the burst temporarily enhanced the accretion rate, a phenomenon observed in other superburst sources, such as 4U~1820--30. \cite{Ballantyne_2004} showed that the inner region of the accretion disk appeared to be disrupted and then reformed over a timescale of about 1000 seconds following a superburst. They interpreted changes in iron K$\alpha$ line and edge features as evidence for this evolution. \cite{Keek_2014,2014ApJ...789..121K} studied a superburst from 4U~1636--536 and also found evidence for reflection features like the iron line and edge parameters evolving over the burst duration, indicating a changing ionization state. They also noted an increase in the persistent flux during the superburst, suggesting enhanced accretion. However, these variations are often observed within a day or so of the superburst. This is the first time we see the effect of a superburst towards the accretion flow at such long time scales. Changes in the accretion disk properties might be observed on shorter timescales in the case of Ultra-compact X-ray binaries~(UCXBs). Depending on the impact, the accretion emission could become suppressed, as seen during the superburst from 4U~1820--30 \citep{2025arXiv250407329J} and the intermediate-duration burst from IGR~J17062--6143 \citep{2021ApJ...920...59B}. It is intriguing to observe the increase in flux around MJD~59047, shortly after the 2020 superburst. This could indicate that the accretion disk, if disrupted, was in the process of being restored and stabilizing. A similar trend might be observed in the NICER light curve, potentially showing a rise followed by stabilization. 
The long time scale evolution of the inner disk radius displayed in the right panel of \autoref{fig:decay_mod} suggests the impact of nonviscous processes. An $\alpha$ disk \citep{Shakura1973A&A....24..337S} has at radius $r$ a viscous timescale $t_\nu=r^2/\nu$, with $\nu$ being 
\begin{equation}\begin{split}
    \nu = &\left[3.6\times10^{17}\,\mathrm{cm}^2\,\mathrm{s}^{-1}\right] \left(1-\sqrt{\frac{R_*}{r}}\right)^{2}\left(\frac{\dot{M}}{\dot{M}_{Edd}}\right)^2 \\
    &\times\left(\frac{R}{r_s}\right)^{-3/2}  \alpha \left(\frac{\eta}{0.1}\right)^{-2} \left(\frac{ M  }{M_\odot}\right)\end{split}\label{eq:nu_rad}
\end{equation}
in the radiation-pressure dominated regime, and 
\begin{equation}\begin{split}
    \nu = &\left[1.5\times10^{12}\,\mathrm{cm}^2\,\mathrm{s}^{-1}\right]\left(1-\sqrt{\frac{R_*}{r}}\right)^{2/5}\alpha^{4/5} \left(\frac{\dot{M}}{\dot{M}_{Edd}}\right)^{2/5}  \\
    &\times\left(\frac{R}{r_s}\right)^{3/5} \left(\frac{M}{M_\odot}\right)^{4/5}\left(\frac{\eta}{0.1}\right)^{-2/5}\end{split}\label{eq:nu_gas}
\end{equation} 
in the gas-pressure dominated regime. The radius $r_s$ is the Schwarzschild radius, defined as $2GM/c^2$, with $G$ being the gravitational constant and $c$ the speed of light. Setting the neutron star mass $M$ to $1.57 M_\odot$ and radius $R_*$ to 9.8 km \citep{2016ApJ...820...28O}, the mass accretion rate $\dot{M}$ to $1.76\times10^{17}$ g s$^{-1}$  \citep{Bhattacherjee2024ApJ...971..154B}, the Eddington mass accretion rate $\dot{M}_{Edd}$ to $\approx2.2\times10^{18}$ g s$^{-1}$ assuming pure hydrogen accretion, and the radiative efficiency $\eta$ to $0.1$, allows solving for $r$ for a given viscosity parameter $\alpha$ and time scale. The black solid lines in \autoref{fig:rad_disk_sim} trace the radii for $t_\nu=1.18$ days in the radiation-pressure dominated (left panel) and the gas-pressure dominated regime (right panel). The grey shaded areas account for the $0.35$ day uncertainty in $t_\nu$. The small obtained inner disk radii (right panel of \autoref{fig:decay_mod}) and the long $t_\nu$ time scale would require $\alpha\lesssim 10^{-6}$ for either disk regime. Given that such small viscosity parameters have not been obtained in other studies imply that $t_\nu$ does not adequately capture the disk evolution, and that the long time scale evolution after the superburst is not primarily driven by viscous processes \citep[see, ][]{2007MNRAS.376.1740K}.

The radial evolution in \autoref{fig:decay_mod} yielding the timescale of 1.18 days disagrees with theoretical expectations and may be connected to the temperature evolution in the accretion disk or to coronal changes instead. X-ray bursts are expected to exert Poynting-Robertson \citep[][]{Robertson1937MNRAS} drag on the disk material, which drains the inner disk region \citep{Walker1992ApJ}. Thus, the inner disk radius is expected to move outward during the burst, which has also been observed in simulations \citep{Fragile2020NatAs,Speicher2023MNRAS}. The observationally inferred inner disk radius depends on the color-correction factor (equation \ref{eq:rin}), which is temperature dependent \citep[e.g.,][]{Zdziarski2022ApJ}. The temperature of simulated accretion disks decreases more slowly than the burst light curve in the tail \citep{Speicher2024ApJ}, suggesting that the accretion disk temperatures could still be above preburst levels after the burst. The higher associated color-correction factor would decrease the inner disk radius estimate. Alternatively, the inner disk radius could be underestimated due to some changes in the corona that affect how much energy is dissipated there \citep{Merloni2000MNRAS}. However, current simulations do not capture the coronal evolution in the burst tail \citep{Fragile2018ApJ}, so it is unclear on what timescales the corona would recover from the burst. Nevertheless, it is plausible that the inner disk radius estimates in \autoref{fig:decay_mod} do not represent the physical radii, which could also explain why they do not evolve on the viscous timescale.

\section{Summary and Conclusions}
\label{sec:conclusions}

We present evidence for the brightest superburst detected from \src by MAXI, at around 00:45 UTC on 16 July 2020. We compare the outbursts and the superbursts observed from the source to understand the conditions causing superbursts. Finally, using pointed observations of \src, we investigate possible effects of the 2020 superburst on the accretion flow. Our main conclusions are as follows.

\begin{itemize}

\item In Section \ref{sec:comparison1}, we compare the outbursts during which a superburst has been observed from \src, and show that only during the brightest outbursts a superburst has been detected. Given the limitation in coverage, this result may not be conclusive and the conditions for a superburst to happen may not only depend on one parameter alone. \cite{Keek_2008} suggest much longer waiting times for superbursts to occur and at least for \src we can see that the time between superburst can be as short as 5-15 years. However, our results indicate that brighter outbursts may at least act as a tipping point for the conditions for a superburst to be reached.

\item We compare the three superbursts observed from \src and show that the 2020 event reached a much higher peak flux compared to the 2025 event and the 2020 superburst likely had the highest ignition column depth among all three events. 

\item Detailed spectral analysis of the pointed observations of \src before and after the 2020 superburst shows that the accretion flow was affected by the superburst for a period that lasts at least a few days. In \autoref{fig:spec_ev}, we observe a sudden increase in the inner disk temperature and a decrease in the inner disk radius, followed by a gradual decrease/recovery over approximately one day. This temporal evolution suggests a sustained influence of the burst radiation on the inner accretion flow.

\item The inferred viscous timescale requires an unrealistically low viscosity parameter ($\alpha\lesssim 10^{-6}$), therefore we conclude that viscous processes alone are not enough to explain the observed disk evolution timescale. These findings imply that the disk evolution after the superburst is not driven solely by viscous processes~\citep[see e.g.,][]{2007MNRAS.376.1740K}.

\item The observed evolution of the inner disk radius after superburst, with a timescale of 1.18 days, deviates from theoretical viscous timescale expectations. This inconsistency may arise from temperature-dependent changes in the disk or corona, suggesting that the inferred radius values might not reflect true physical changes but rather variations in spectral modeling parameters i.e., color correction factor.

\end{itemize}

\section{Acknowledgements}

We would like to thank our referee for helpful and valuable comments on the manuscript. We would also like to thank Dr. Montserrat Armas Padilla for her helpful explanations and discussions regarding her paper and the use of the NTHCOMP model.
 T.G. greatly appreciates the hospitality shown at the Georgia Institute of Technology School of Physics, where part of this work is completed. T.G., T.B., E.E.D., are supported in part by the Turkish Republic, Directorate of Presidential Strategy and Budget project, 2016K121370. M.N. is a Fonds de Recherche du Québec – Nature et Technologies (FRQNT) postdoctoral fellow. This work was supported by NASA through the \nicer mission and the Astrophysics Explorers Program. This research has made use of the MAXI data provided by RIKEN, JAXA and the MAXI team~\citep{Matsuoka_2009}. This research has made use of data and/or software provided by the High Energy Astrophysics Science Archive Research Center (HEASARC), which is a service of the Astrophysics Science Division at NASA/GSFC.
 
\section*{Data Availability}
All the data used in this publication are publicly available through NASA/HEASARC, Insight-HXMT, and MAXI data archives.


\bibliographystyle{mnras}
\bibliography{example} 



\bsp	
\label{lastpage}
\end{document}